\documentstyle[11pt,aaspp4]{article}

\input{psfig.sty}

\newcommand\sgra {SGR~1806$-$20}

\newcommand\sgrc {SGR~1900$+$14}
\newcommand\sgrd {SGR~1627$-$41}

\newcommand\axpa {1E~2259$+$586}
\newcommand\axpb {1E~1048.1$-$5937}

\newcommand\axpf {XTE~J1810$-$197}
\newcommand\axpg {AX~J1845$-$0258}              

\newcommand\psr {PSR~J1811$-$1925}

\newcommand\rxte {{\it RXTE}}

\begin{document}

\title{X-ray Bursts from the Transient Magnetar Candidate  XTE~J1810$-$197}

\author{
Peter~M.~Woods\altaffilmark{1,2},
Chryssa~Kouveliotou\altaffilmark{3,2},
Fotis~P.~Gavriil\altaffilmark{4},
Victoria~M.~Kaspi\altaffilmark{4,5}, 
Mallory~S.E.~Roberts\altaffilmark{4}, 
Alaa~Ibrahim\altaffilmark{6},
Craig~B.~Markwardt\altaffilmark{7},
Jean~H.~Swank\altaffilmark{7}, and
Mark~H.~Finger\altaffilmark{1,2}
}

\altaffiltext{1}{Universities Space Research Association; 
peter.woods@nsstc.nasa.gov}
\altaffiltext{2}{National Space Science and Technology Center, 320 Sparkman Dr. 
Huntsville, AL 35805}
\altaffiltext{3}{NASA at Marshall Space Flight Center}
\altaffiltext{4}{Department of Physics, Rutherford Physics Building, McGill 
University, 3600 University St., Montreal, Quebec, H3A~2T8, Canada}
\altaffiltext{5}{Canada Research Chair and Steacie Fellow}
\altaffiltext{6}{Deprartment of Physics, George Washington University, 725
21$^{\rm st}$ Street, NW, Washington, DC 20052}
\altaffiltext{7}{NASA at Goddard Space Flight Center, LHEA, Mail Code 662,
Greenbelt, MD 20771}

\begin{abstract}

We have discovered four X-ray bursts, recorded with the {\it Rossi X-ray Timing
Explorer} Proportional Counter Array between 2003 September and 2004 April,
that we show to originate from the transient magnetar candidate \axpf.  The
burst morphologies consist of a short spike or multiple spikes lasting $\sim$1
s each followed by extended tails of emission where the pulsed flux from \axpf\
is significantly higher.  The burst spikes are likely correlated with the pulse
maxima, having a chance probability of a random phase distribution of 0.4\%. 
The burst spectra are best fit to a blackbody with temperatures 4$-$8 keV,
considerably harder than the persistent X-ray emission.  During the X-ray tails
following these bursts, the temperature rapidly cools as the flux declines,
maintaining a constant emitting radius after the initial burst peak.  During
the brightest X-ray tail, we detect a narrow emission line at 12.6 keV with an
equivalent width of 1.4 keV and a probability of chance occurrence less than 4
$\times$ 10$^{-6}$.  The temporal and spectral characteristics of these bursts
closely resemble the bursts seen from \axpb\ and a subset of the bursts
detected from \axpa, thus establishing \axpf\ as a magnetar candidate.  The
bursts detected from these three objects are sufficiently similar to one
another, yet significantly different from those seen from soft gamma repeaters,
that they likely represent a new class of bursts from magnetar candidates
exclusive (thus far) to the anomalous X-ray pulsar-like sources.

\end{abstract}


\newpage

\section{Introduction}

Originally grouped into two distinct, empirically-defined classes, Anomalous
X-ray Pulsars (AXPs) and Soft Gamma Repeaters (SGRs) are now widely recognized
as fundamentally the same type of object.  Historically, some of the salient
properties of AXPs include spin periods within a narrow range (5$-$12 s), rapid
spin-down rates ($10^{-13}-10^{-10}$ s s$^{-1}$), X-ray luminosities within the
range $\sim$10$^{33}-10^{36}$ ergs s$^{-1}$, soft energy spectra most often
characterized by the sum of a power-law and blackbody, and very dim optical/IR
counterparts.  The defining property of SGRs was their emission of brief
($\sim$0.1 s), luminous ($\gtrsim$10$^{38}$ ergs s$^{-1}$) X-ray bursts.  Two
notable similarities drawn between AXPs and SGRs in recent years that have in
large part established the connection between the two classes were the
discovery of X-ray pulsations and rapid spin down in two SGRs (Kouveliotou et
al.\ 1998, 1999; Hurley et al.\ 1999), a defining AXP trait, and the discovery
of bright X-ray bursts from two AXPs (Gavriil, Kaspi \& Woods 2002; Kaspi et
al.\ 2003), a defining SGR trait.  These two classes of isolated neutron stars
have each been identified as probable manifestations of highly magnetized
neutron stars (i.e.\ magnetars) whose bright X-ray emission is powered by the
decay of their strong magnetic fields of the order of 10$^{15}$ Gauss at the
stellar surface (Thompson \& Duncan 1995, 1996).  Hereafter, we will refer to
AXPs and SGRs collectively as magnetar candidates.  For a recent review of
these objects, see Woods \& Thompson (2004).

The most recent addition to the small (11 confirmed), but growing class of
magnetar candidates, \axpf, was discovered in 2003 July (Ibrahim et al.\ 2004),
interestingly enough, during a target-of-opportunity {\it Rossi X-ray Timing
Explorer} (\rxte) observation of another magnetar candidate nearby on the sky
(\sgra).  The X-ray emission from \axpf\ was found to be spectrally soft and
pulsed with a period of 5.54 s.  Additional \rxte\ observations showed that the
flux from \axpf\ was decaying, the pulsar was spinning down rapidly, and there
was no evidence for orbital Doppler shifts due to a binary companion (Markwardt
et al.\ 2003).  Based upon these characteristics, the new X-ray source was
tentatively identified as a transient AXP.

A subsequent ToO observation of \axpf\ with {\it XMM-Newton} showed that the
source had a typical AXP energy spectrum with a blackbody temperature of 0.7
keV and a power-law photon index $-$3.7 (Gotthelf et al.\ 2004).  The location
provided by a ToO {\it Chandra} observation showed that this source has been
present in archival X-ray observations dating back to 1980, albeit at a flux
level some two orders of magnitude dimmer and with a softer energy spectrum
(Gotthelf et al.\ 2004). This level of variability is extreme within the class
of magnetar candidates, but not unprecedented (e.g.\ \sgrd\ [Kouveliotou et
al.\ 2003] and the AXP candidate \axpg\ [Vasisht et al.\ 2000]).  The
sub-arcsecond {\it Chandra} location also led to the discovery of the likely IR
candidate to \axpf\ with properties closely resembling those of the other known
AXPs (Israel et al.\ 2004).

Here, we report the definitive evidence for the identification of \axpf\ as a
magnetar candidate.  We have discovered four X-ray bursts from the source
recorded during \rxte\ observations of this region.  We report on the temporal
and spectral properties of these bursts and show that they most closely
resemble bursts seen from other magnetar candidates, in particular, \axpb\
(Gavriil et al.\ 2002, 2005) and \axpa\ (Gavriil et al.\ 2004).  Finally, we
discuss the possibility that a new class of magnetar candidate bursts is
emerging whose characteristics include long durations, a temporal correlation
of the bursts with the peak of the pulsed X-ray emission, and a spectral
hardness versus burst intensity correlation.

\section{\rxte\ Observations}

We have been monitoring \axpf\ with \rxte\ since 2003 through a combination of
dedicated observations pointed directly at the source and observations of two
nearby objects.  The two nearby X-ray sources monitored separately with \rxte,
\sgra\ and \psr, are 0.74 and 0.49 degrees offset in angle from the transient
magnetar candidate, well within the 1.1 degree radius (zero response)
Proportional Counter Array (PCA [Jahoda et al.\ 1996]) field-of-view (FOV). 
Here, we report on our burst search through PCA data from all \rxte\
observations containing \axpf\ between 2003 January and 2004 September.  Over
this time interval, the source was observed numerous times for a total exposure
of 791 ks.  Analysis of the persistent and pulsed emission from \axpf\ for a
subset of these data is presented in Ibrahim et al.\ (2004).  A detailed
analysis of the full data set will be presented elsewhere.

\section{X-ray Bursts from \axpf}

\begin{figure}[!htb]

\centerline{
\psfig{file=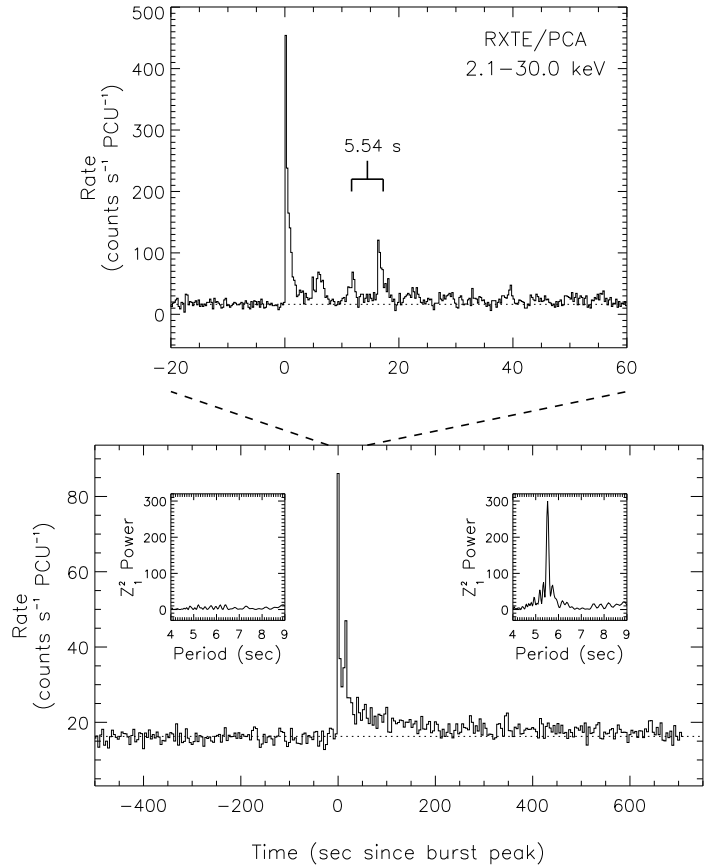,height=5.1in}}
\vspace{-0.1in}

\caption{Time history of the 2004 February 16 burst (Burst 2) from \axpf\
recorded with the {\it RXTE} PCA.  {\it Top panel} -- A close-up of the
short-duration, bright initial peak with 0.25 s time resolution.  Note the 5.54
s pulsations following the main peak.  {\it Bottom panel} -- An expanded view
of the burst with 4 s time resolution showing the long X-ray tail following the
initial spike.  The total pre-burst flux level is denoted by the horizontal
dashed line.  The source becomes occulted by the Earth at T$+$706 s, before the
flux returns to its pre-burst level.  Inset panels show the Z$^2_1$ power
spectra for 200 s of data before ({\it left}) and after ({\it right}) the burst
peak. \label{fig:burst2_lc_z2}}

\end{figure}

While filtering out bursts from a ToO observation of \sgra\ carried out on 2004
February 16, we noted that one of the 11 bursts detected during this
observation lasted $\sim$2 s and was followed by an X-ray tail lasting several
hundred seconds (Figure~\ref{fig:burst2_lc_z2}).  X-ray tails or burst
afterglows have been detected in several magnetar candidates (e.g.\ Woods et
al.\ 2004), so, in principle, this was not unexpected behavior from the SGR. 
In fact, we have detected X-ray tails following two separate bursts from \sgra\
(Woods et al.\ in preparation) whose associations with that source were
confirmed by Inter-Planetary Network (IPN) locations.  However, an unusual
property of the burst/afterglow pair observed on 2004 February 16 was the
relatively small fluence of the burst compared with the fluence of the
afterglow (1:6 in terms of total counts).  The other two afterglows with bursts
{\it confirmed} to have come from \sgra\ had burst-to-afterglow fluence ratios
several orders of magnitude higher.  Moreover, a much brighter burst recorded
earlier during the 2004 February 16 observation, presumably from \sgra, showed
no evidence for an X-ray afterglow.  We concluded that either this was a new
type of burst never before seen in \sgra, or it was a burst from a different
source in the PCA FOV.

Knowing that \axpf\ was within the PCA FOV and that magnetar candidates have
always shown an enhanced pulsed flux amplitude during X-ray tails following
bursts (e.g.\ Lenters et al.\ 2003), we constructed a $Z^2_1$ power spectrum
before and after the burst (both \sgra\ and \axpf\ have very simple pulse
profiles).  For 200 s exposures directly before and after the burst, we find
that the pulsed amplitude of \sgra\ (7.56 s) does not change significantly
while the \axpf\ pulsed amplitude (5.54 s) increases dramatically
(Figure~\ref{fig:burst2_lc_z2} - Burst 2).  The increase in pulsed amplitude
was so large that one can clearly identify individual pulsations in the light
curve.  Due to the profound pulsed flux increase of \axpf\ simultaneous with
this burst, we conclude that it was emitted from \axpf.

\vspace{-0.25in}

\begin{figure}[!htb]

\centerline{
\psfig{file=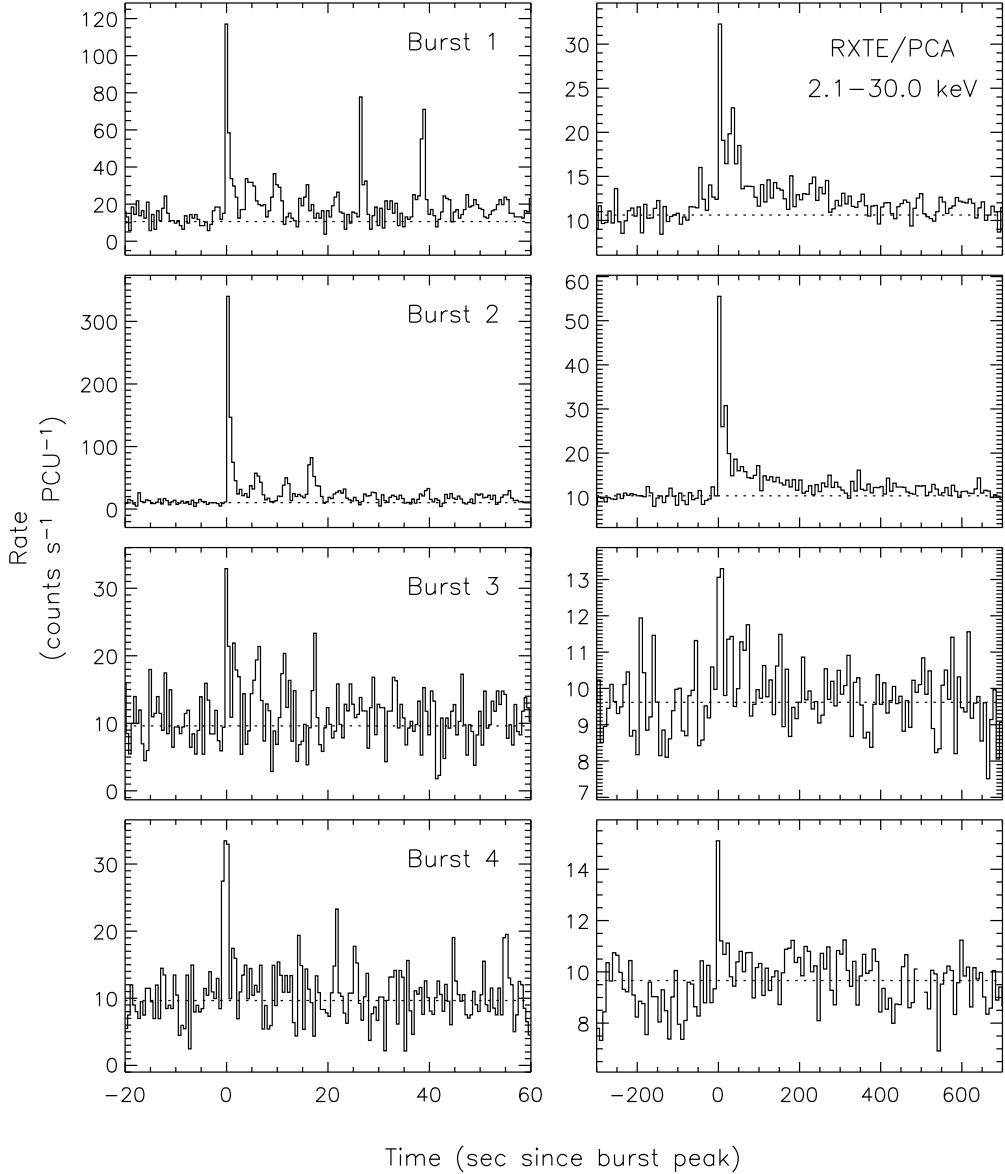,height=7.0in}}

\caption{Time histories of the four bursts recorded from \axpf\ with the {\it
RXTE} PCA (2.1$-$30 keV).  The four left panels on show each burst with 0.5 s
time resolution.  Right panels show the same bursts over an expanded time range
with 8 s time resolution.  The horizontal dotted lines indicate the total
pre-burst flux level.  For burst times, durations, etc., see
Table~\ref{tab:bursts}. \label{fig:all_bursts}}

\end{figure}

Motivated by this discovery, we searched all \rxte\ PCA observations containing
\axpf\ for similar bursts.  In our search, we first identified bursts within
2$-$30 keV light curves for each observation.  We defined a burst as a
$>5\sigma$ excess above background on the 0.25 s time scale where the counts
were consistent with being equally distributed among all active PCUs.  The
background rate during each bin was calculated by taking an average of 30 s
before and after the bin being searched.  Next, we measured the pulsed
amplitude of \axpf\ for short integrations (30 cycles or 166.2 s) throughout
the observation.  Four bursts, including the one described above, showed
simultaneous pulsed flux excesses (Figure~\ref{fig:all_bursts}).  Only one
observation (80150-01-05-00) showed a marginal pulsed flux enhancement without
any corresponding burst.  However, the pulsed flux decreased monotonically
through the first half of the 2.5 ks observation which may be indicative of an
undetected burst immediately preceeding this observation.  Several hundred
bursts were detected without corresponding pulsed flux increases, but only
during observations centered on \sgra.  At this point, we cannot distinguish
between \sgra\ and \axpf\ as the origin of the vast majority of these bursts,
although most are likely to have come from \sgra.  Several tens of these bursts
were detected and localized by INTEGRAL and/or the IPN, and all of those
reported thus far have been confirmed to originate with \sgra.

We constructed $Z^2_1$ power spectra for 200 s of data before and after each of
the four bursts and found that the pulsed amplitude increase occurred {\it
following} the burst in all cases.  The change in 2$-$30 keV pulsed amplitude
($\Delta A$) across each burst is listed in Table~\ref{tab:bursts}.  For Bursts
1 and 2, the increase is highly significant ($>9\sigma$).  For the two weaker
bursts (3 and 4), the change in pulsed amplitude is marginal ($3.5\sigma$ and
$2.4\sigma$, respectively).  However, the rate increases of the bursts
themselves are not marginal ($>5.7\sigma$) and these two bursts, in addition to
Burst 1, were recorded during observations of \psr.  These pointings exclude
\sgra\ from the PCA FOV (1.2 degrees off-axis), thus eliminating this object as
the origin of Bursts 1, 3 and 4.  We, therefore, attribute all four bursts to
\axpf.

\section{Temporal Analysis}

The burst morphologies consist of brief spikes ($\sim$0.5$-$2 s) superposed
upon extended tails of emission strongly modulated at the \axpf\ spin
frequency.  For each of the four bursts, we measured several temporal
parameters including durations, rise times, and phase occurrence of the sharp
spikes within the burst light curves.

\subsection{Durations}

First, we measured $T_{90}$ durations, defined as the time over which 90\% of
the burst counts are accumulated (Kouveliotou et al.\ 1993; Koshut et al.\
1996), for each of the four events.  Due to the quasi-exponential flux decay
following these bursts, we first subtracted the background due to the passage
of \rxte\ through Earth's radiation belts using {\tt PCABACKEST}.  Next, we fit
a low-order polynomial through several hundred seconds of data preceding and
following each burst, well away from intervals containing burst emission.  We
interpolated this polynomial model through the burst interval to subtract the
remaining background, then measured $T_{90}$ durations (2$-$30 keV) for each
event.  While the bright spikes themselves are relatively brief, the total
burst durations are much longer (Table~\ref{tab:bursts}).  Note that only a
lower limit could be determined for Burst 2 due to Earth-occultation of \axpf\
during this event.

\subsection{Burst Spikes versus Pulses}

At times, the distinction between bright pulses and bursts was not obvious.  We
set out to separate ordinary pulses from what we will refer to as ``burst
spikes'' within the time history of each event in the following way.  From 70 s
prior to the peak of each burst until 200 s after, we split the time history
into equal length segments of one pulse period (5.54 s).  We binned the
recorded counts within each segment into 16 phase bins and constructed a
Fourier representation of each pulse using 5 harmonics.  The Fourier
representations of the individual pulse profiles were cross-correlated with the
average pulse profile derived from a much longer time integration of the
persistent emission.  For each 5.54 s integration, we fit for both the pulse
phase and amplitude and calculated the goodness-of-fit ($\chi^2$) which is
defined as

\begin{equation}
\chi^2 = \sum_{k=1}^{5} \frac{|\alpha_{k} - 
   I \tau_{k} e^{2 \pi i k \Delta \phi}|^2}{\sigma^{2}_{k}},
\end{equation}

\noindent where 

\begin{eqnarray*}
\alpha_{k} = \frac{1}{N} \sum_{j=1}^{N} r_j e^{-2{\pi}i\phi_j{k}}, 
\hspace{1.0cm}
\tau_{k} = \frac{1}{N} \sum_{j=1}^{N} p_j e^{-2{\pi}i\phi_j{k}}, 
\hspace{1.0cm}
\sigma_{k}^{2} = \frac{1}{2N^2} \sum_{j=1}^{N} \sigma^{2}_{r_j}. \\
\end{eqnarray*}

\begin{figure}[!htb]

\centerline{
\psfig{file=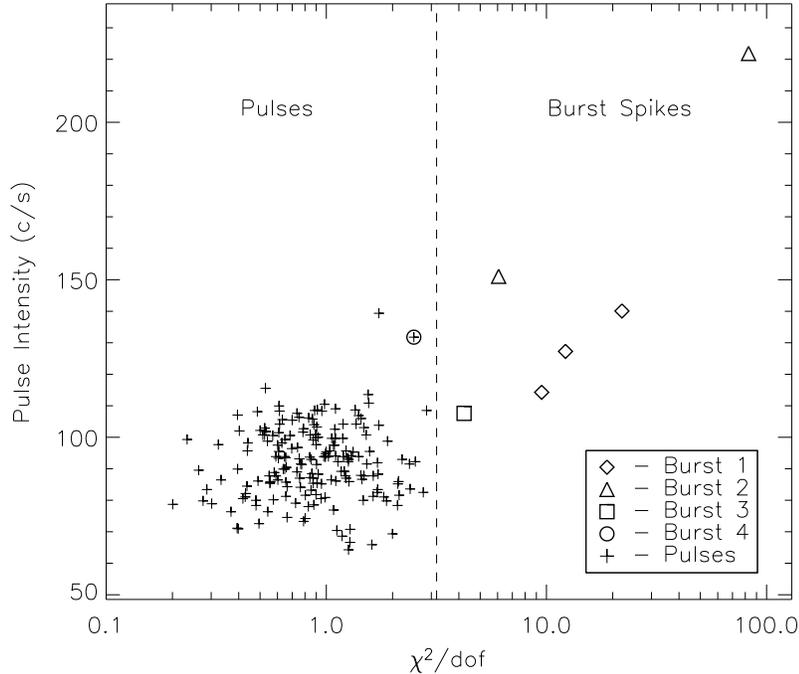,height=3.9in}}
\vspace{-0.05in}

\caption{Pulse intensity versus $\chi^2$/dof for individual pulses recorded
during the four bursts observed from \axpf.  The $\chi^2$ parameter measures
the morphological consistency with the pulse profile of the persistent
emission.  The vertical dashed line marks the boundary between ordinary pulses
and what we refer to as burst spikes.  See text for details.
\label{fig:peak_pulse}}

\end{figure}

\noindent Here, $k$ refers to the harmonic number, $j$ refers to the phase bin,
$N$ is the total number of phase bins, $\phi_j$ is the phase, $p_j$ is the
count rate in the $j^{\rm th}$ phase bin of the template, $r_j$ is the count
rate in the $j^{\rm th}$ phase bin of the individual pulse, $\sigma_{r_j}$ is
the uncertainty in the count rate of the $j^{\rm th}$ phase bin of the
individual pulse, $I$ is the pulsed amplitude, and $\Delta \phi$ is the phase
shift.  The number of degrees of freedom for the fit (8) is twice the number of
harmonics used in the Fourier representation (5) minus the number of free
parameters ($I$ and $\Delta \phi$).  Ordinary pulses returned low $\chi^2$
values confirming their morphological consistency with the pulse shape. The
burst spikes, on the other hand, yielded high $\chi^2$ values and high pulsed
intensities (Figure~\ref{fig:peak_pulse}).

We set a threshold of 3$\sigma$ ($\chi^2$ = 25.4 for 8 dof) to define the
boundary between burst spikes and pulses.  We identified 3, 2, 1 and 0 burst
spikes in Bursts 1, 2, 3 and 4, respectively.  Note that the initial pulse of
Burst 4 ($\chi^2_{\nu}$ = 2.5) falls slightly below our threshold, therefore
this entire event was categorized, following our definition, as a sequence of
bright pulses with no corresponding burst spike.  Peak times of the burst
spikes and the initial bright pulse of Burst 4, for completeness, are listed in
Table~\ref{tab:bursts}.  The average $\chi^2_{\nu}$ (i.e.\ reduced $\chi^2$)
for all individual pulses was 1.00$\pm$0.04 where the error in the mean is
determined from the sample variance.  Thus, the individual pulses from \axpf\
seen during these burst tails, even several very high amplitude pulses, are
morphologically consistent with the average pulse profile.  Although several
pulses have amplitudes higher than a few of the burst spikes, there is still a
selection effect against burst spikes with low amplitudes due to low count
statistics.

\subsection{Burst Spike Morphologies}

\begin{figure}[!htb]

\centerline{
\psfig{file=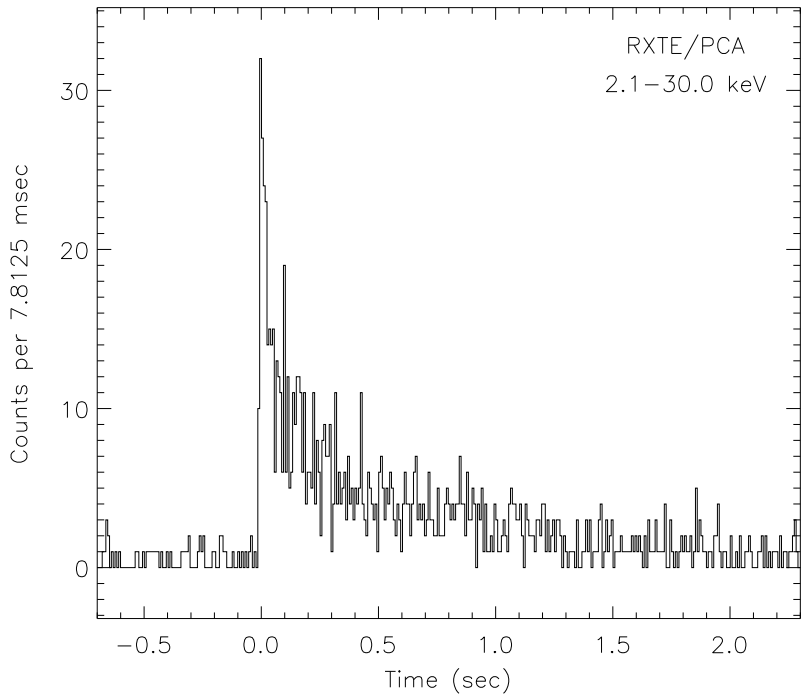,height=4.0in}}

\caption{Time history of the initial peak of the 2004 February 16 burst (Burst
2) from \axpf\ recorded with the {\it RXTE} PCA.  Note the rapid rise and
quasi-exponential decay. \label{fig:burst2_peak1_zoom}}

\end{figure}

The individual burst spikes range in duration between $\sim$0.5 s and 2 s,
although very precise determinations of durations are not possible due to the
inseparable, underlying pulsations.  The morphologies of the burst spikes are
each consistent with a faster rise time than decay time. We measured the rise
times of the burst spikes by fitting the local time histories ($\pm$2.5 s) to a
model consisting of a linear rise followed by an exponential decay.  This model
has been used previously to fit magnetar candidate burst time histories
(Gavriil et al.\ 2002), and it adequately describes the burst morphologies
observed here.  For example, a close-up of the brightest observed burst spike
(spike 1 of Burst 2) is shown in Figure~\ref{fig:burst2_peak1_zoom}.  In
general, the peak of each burst spike is defined as the time of the highest
count density within the burst interval.  To avoid statistical fluctuations,
the maximum count density is determined for six consecutive counts where the
mean arrival time of these events defines the burst peak.  Due to the small
number of total burst counts, we fit the model to the individual counts within
the burst interval using a maximum likelihood method.  The rise time is defined
as the time between the intersection of the linear term with the pre-burst
background count rate and the burst peak.  The rise times of these burst spikes
range from 14 ms to 150 ms (Table~\ref{tab:bursts}).  The peak of Burst 4,
classified earlier as a somewhat bright, yet ordinary pulse, has a much slower
rise time of $\sim$600 ms.

\subsection{Burst Tail Pulse Properties}

\begin{figure}[!htb]

\vspace{-0.3in}

\centerline{
\psfig{file=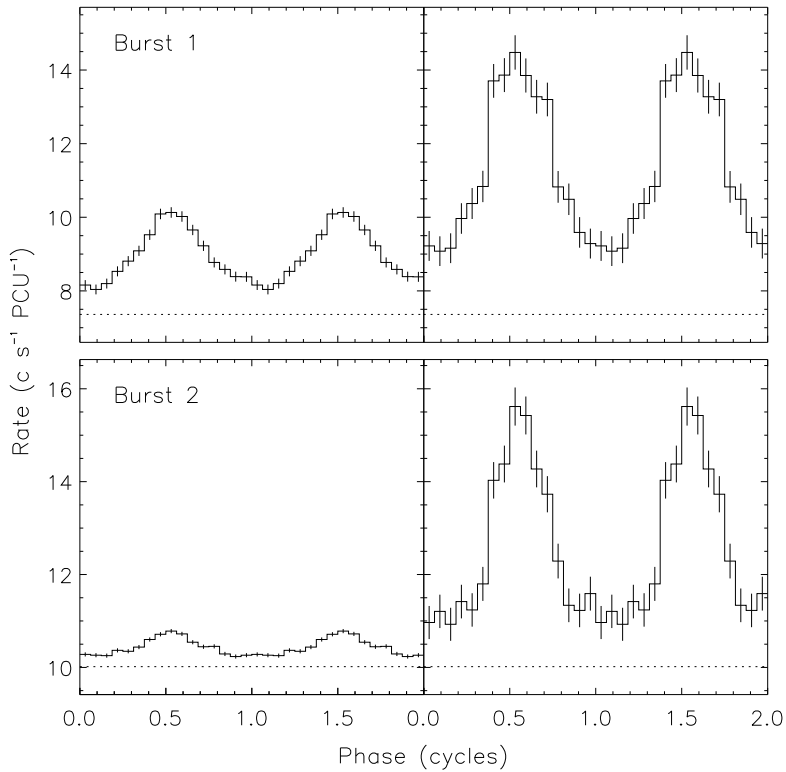,height=4.5in}}

\caption{Folded pulse profiles of \axpf\ just before ({\it left panels}) and
during the tails of ({\it right panels}) the two brightests bursts (Burst 1 and
Burst 2).  All pulse profiles have 16 bins per cycle and were constructed from
counts recorded with the {\it RXTE} PCA within the energy range 2$-$10 keV. 
The horizontal dotted lines in all four panels denote the pre-burst background
level.  See text for details on how this background level was determined and
the measured pulsed fractions during the tails. \label{fig:tail_profile}}

\end{figure}

The individual pulses within the tails of these bursts have morphologies
consistent with the average pulse profile seen in the persistent X-ray emission
(Figure~\ref{fig:peak_pulse}), although this comparison is limited by the small
number of counts recorded within individual pulses.  Here, we probe the pulse
morphology during the two brightest burst tails with higher sensitivity and
investigate the phase alignment of these pulses with the pre-burst pulsations.
First, we folded the 2$-$10 keV events recorded within the $T_{90}$ intervals
of Burst 1 and 2, excluding the burst spikes, on the spin ephemeris of \axpf. 
The resulting burst tail pulse profiles are well aligned with the pre-burst
pulse profiles and have similar morphologies (Figure~\ref{fig:tail_profile}). 
We find a 3$\sigma$ upper limit of 0.03 cycles on a phase shift during the
tails of these two bursts.  The average pulse shape during the burst tails was
slightly altered relative to the pre-burst pulsations.  Specifically, the pulse
peak during the tails is narrower than the pre-burst pulse peak.

Motivated by the modest changes in pulse shape that were measured during these
two burst tails, we investigated the pulsed fraction following the bursts.  It
is not possible to accurately measure the pulsed fraction of the {\it
persistent} emission due to the non-imaging PCA instrument, low count rate of
\axpf\ at these epochs, and uncertain contribution to the PCA background due to
other sources in the Galactic plane.  We can, however, measure the pulsed
fraction during the burst {\it tails} following Lenters et al.\ (2003).  First,
we subtracted the ``local'' PCA background using {\tt PCABACKEST}.  Next, we
estimate the cosmic background (assumed to be constant within the $\sim$10 ks
observations) by varying the background count rate until the pre-burst r.m.s.\
pulsed fraction equaled 42.7$\pm$0.8\% and 41.3$\pm$0.8\%, for Bursts 1 and 2,
respectively.  These 2$-$10 keV pulsed fractions were calculated from a linear
interpolation between pulsed fraction measurements obtained from four {\it
XMM-Newton} observations of \axpf\ (manuscript in preparation).  We then used
this rate as our background level for the burst tail (shown as horizontal
dotted lines in Figure~\ref{fig:tail_profile}) and measured r.m.s.\ pulsed
fractions (2$-$10 keV) of 48.0$\pm$3.1\% and 61.5$\pm$4.3\% during the tails of
the respective bursts.  The pulsed fraction during the tail of Burst 1 is
consistent with the pre-burst level, whereas Burst 2 shows a significant
increase in pulsed fraction.  We conclude that the beaming function responsible
for the persistent emission pulsations changes significantly during the tail of
the brightest burst from \axpf, such that the pulse maximum increases by a
larger factor than pulse minimum.  The pulsed fraction increase and pulse
morphology change during this burst tail could be caused by unresolved bursts
superposed upon pulse peaks.

\subsection{Burst Phase Occurrence}

\begin{figure}[!htb]

\centerline{
\psfig{file=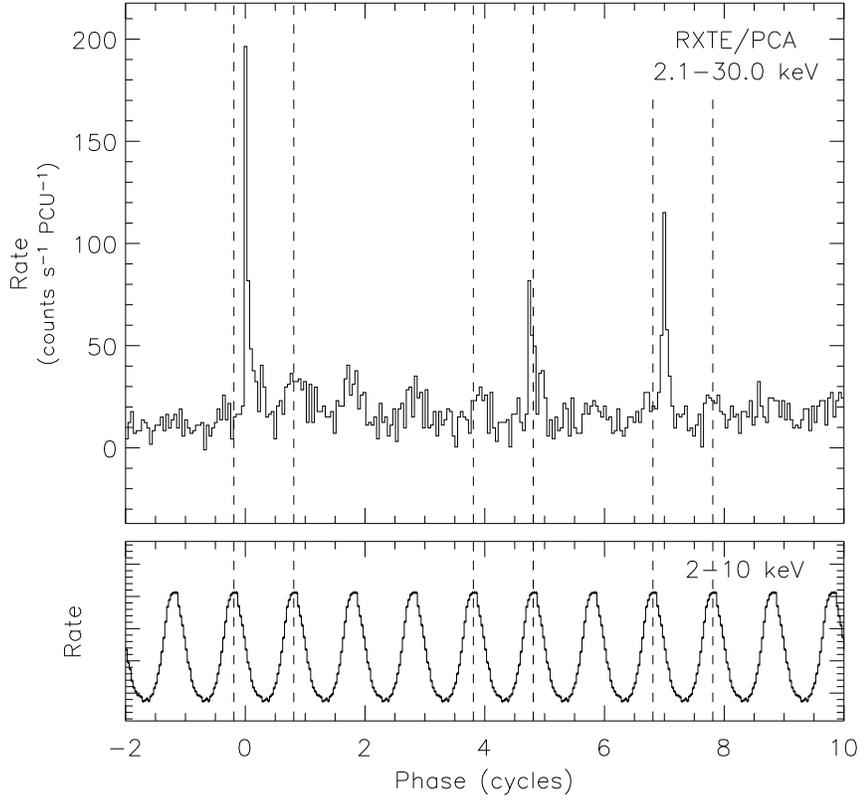,height=4.7in}}

\caption{{\it Top panel} -- Light curve of the 2003 September 22 burst (Burst
1) from \axpf\ recorded with the {\it RXTE} PCA.  Time has been converted to
pulse phase to compare the burst peaks with the pulse profile.  {\it Bottom
panel} -- The folded pulse profile of \axpf\ repeated multiple times.  The
vertical dashed lines in each panel denote the centroid of pulse maximum.  Note
the close relation of the three primary burst peaks to pulse maximum.
\label{fig:burst1_phase}}

\end{figure}

From a visual inspection of the underlying pulsations within the burst time
histories, it appeared that the burst spikes were concentrated near the maxima
of the pulsed X-ray emission.  We measured the relative alignment of each burst
peak with the nearest pulse maximum in the following way.  The centroid of the
pulse maximum was determined by fitting a quadratic to the portion of the
average pulse profile near pulse maximum.  The average pulse profile used here
was constructed from 230 ks of PCA data.  The phase of each burst was
determined by converting the time of the burst peak to pulse phase using the
same pulse ephemerides utilized to generate the average pulse profile. 
Figure~\ref{fig:burst1_phase} shows the light curve of Burst 1 converted to
pulse phase with the average pulse profile repeated below.  The measured phase
offsets for these three burst spikes given in Table~\ref{tab:bursts} are the
differences between the burst peaks and the nearest vertical dashed lines.

The relative alignment of the burst peaks with the pulse maximum is limited by
the strong timing noise present in this pulsar (Ibrahim et al.\ 2004).  Based
upon the phase residuals from our timing solutions, we estimate the 1$\sigma$
error in the relative alignment to be $\sim$0.03 cycles.  For a total of six
burst spikes, we measure a net average phase offset of $-$0.0015 cycles and an
absolute value average offset of 0.118 cycles.  For a random burst distribution
in phase, one would expect an absolute value average offset of 0.25 cycles. 
The probability that our measured offsets are consistent with a random
distribution in phase is $(2 \times 0.2)^6 = 0.4$\%.  Thus, we conclude that
these bursts from \axpf\ are in all likelihood correlated with pulse maximum. 
Furthermore, since the net average phase offset is consistent with zero, there
is no preference for the burst spikes to lag or lead the main pulse.


\begin{table}[!p]
\begin{minipage}{1.0\textwidth}
\begin{small}
\begin{center}
\caption{Measured spectral and temporal parameters for bursts from
XTE~J1810$-$197. \label{tab:bursts}} 
\vspace{10pt}
\begin{tabular}{cccccccc} \hline \hline

                 &        \multicolumn{3}{c}{Burst 1}                    &    \multicolumn{2}{c}{Burst 2}        &     Burst 3       &    Burst 4         \\
                 &     Spike 1     &     Spike 2     &     Spike 3       &      Spike 1         &     Spike 2    &                   &    Main Pulse      \\\hline

 Date            & \multicolumn{3}{c}{09/22/03}                          & \multicolumn{2}{c}{02/16/04}          &    04/19/04       &    05/19/04        \\ 
 MJD             & \multicolumn{3}{c}{52904}                             & \multicolumn{2}{c}{53051}             &     53114         &     53144          \\ 
 Peak Time       &    27381.70     &    27407.92     &    27420.36       &    86101.50          &    86117.73    &    82122.70       &    36516.11        \\ 
 (sod UT)$^{a}$  &                 &                 &                   &                      &                &                   &                    \\
 $\Delta A^{b}$  & \multicolumn{3}{c}{2.23$\pm$0.24}                     & \multicolumn{2}{c}{3.70$\pm$0.23}     &   0.61$\pm$0.18   &  0.46$\pm$0.19     \\
 (c s$^{-1}$ PCU$^{-1}$) &         &                 &                   &                      &                &                   &                    \\
 Duration        & \multicolumn{3}{c}{500$^{+170}_{-110}$}               & \multicolumn{2}{c}{$>$575}            & 70$^{+20}_{-9}$   & 370$^{+50}_{-40}$  \\
 (s)             &                 &                 &                   &                      &                &                   &                    \\
 Rise Time       & 94$^{+12}_{-8}$ & 30$^{+3}_{-13}$ & 126$^{+14}_{-13}$ & 13.8$^{+0.2}_{-0.4}$ & 39$^{+3}_{-8}$ & 152$^{+19}_{-36}$ & 610$^{+110}_{-70}$ \\ 
 (ms)            &                 &                 &                   &                      &                &                   &                    \\
 Burst Phase     &      0.19       &    $-$0.08      &       0.17        &    $-$0.01           &    $-$0.07     &     $-$0.20       &    $-$0.10         \\
 (cycles)$^{c}$  &                 &                 &                   &                      &                &                   &                    \\\hline
 Blackbody $kT$  &   8.0$\pm$1.1   &   3.9$\pm$0.6   &   5.4$\pm$0.8     &    7.2$\pm$0.5       &   5.3$\pm$0.7  &    5.8$\pm$2.0    &    3.9$\pm$0.7     \\
 (keV)           &                 &                 &                   &                      &                &                   &                    \\
 Blackbody Radius$^{d}$ & 0.38$\pm$0.11 & 0.81$\pm$0.24 & 0.70$\pm$0.21  &   1.04$\pm$0.13      &  0.94$\pm$0.24 &    0.33$\pm$0.23  &    0.41$\pm$0.15   \\
 (km)            &                 &                 &                   &                      &                &                   &                    \\
 Fluence         &  4.8$\pm$0.5    &   1.4$\pm$0.2   &   2.2$\pm$0.3     &    38.8$\pm$2.3      &   8.5$\pm$0.9  &    0.5$\pm$0.1    &    0.5$\pm$0.1     \\
 (10$^{-9}$ ergs cm$^{-2}$) &      &                 &                   &                      &                &                   &                    \\
 Peak Flux$^{e}$ &   18.4$\pm$4.0  &   6.3$\pm$1.9   &   7.4$\pm$2.4     &    99$\pm$1          &   15.1$\pm$4.9 &    2.1$\pm$1.2    &    0.7$\pm$0.4     \\
 (10$^{-9}$ ergs cm$^{-2}$ s$^{-1}$) &       &       &                   &                      &                &                   &                    \\

\hline\hline
\end{tabular}
\end{center}
\noindent$^{a}$ Seconds of day in Universal Time. \\
\noindent$^{b}$ $\Delta A$ is the change in 2$-$30 keV pulsed amplitude $A$ 
from pre-burst to post-burst (200 s integration). \\
\noindent$^{c}$ The approximate 1$\sigma$ error for each measurement is 0.03
cycles.  See text for details. \\
\noindent$^{d}$ The distance used to calculate the radii is 5 kpc.  For
comparison, the radius inferred for the persistent thermal emission has been
steadily decreasing from 4 to 2 $d_{5}$ km between MJD 52980 and 53266
(Halpern \& Gotthelf 2005). \\
\noindent$^{e}$ The peak flux is measured on the 62.5 ms time scale.
\end{small}
\end{minipage}\hfill
\end{table}

\section{Spectral Analysis}

\subsection{Burst Spikes}

Following standard recipes for \rxte\ PCA spectral
analysis\footnote{http://heasarc.gsfc.nasa.gov/docs/xte/recipes/cook\_book.html},
we accumulated PCA spectra for each of the six burst spikes and main pulse of
Burst 4, corresponding background intervals, and {\tt PCABACKEST} spectra for
both burst and background intervals.  The {\tt PCABACKEST} spectra  estimate
the ``local'' background in the PCA using various housekeeping parameters. 
This background component changes through the \rxte\ orbit on time scales of
several tens of seconds to hours.  Since our background intervals are hundreds
of seconds offset from our burst intervals, the {\tt PCABACKEST} spectra at the
two epochs do not match.  To correct for this, we first subtracted the {\tt
PCABACKEST} spectra from the observed background spectra, then added back in
the {\tt PCABACKEST} spectra derived for the burst interval.  These modified
background spectra are used in all subsequent spectral fits of the PCA data.

The two brightest burst spikes of Burst 1 and 2 were also clearly detected with
the High-Energy X-ray Timing Experiment (HEXTE [Rothschild et al.\ 1998]), so
we accumulated HEXTE spectra from cluster A of burst and background intervals
for these events.  We approximated the background during the burst interval by
choosing background intervals before and after each burst equidistant from the
burst peak.

\begin{figure}[!htb]

\vspace{-0.2in}

\centerline{
\psfig{file=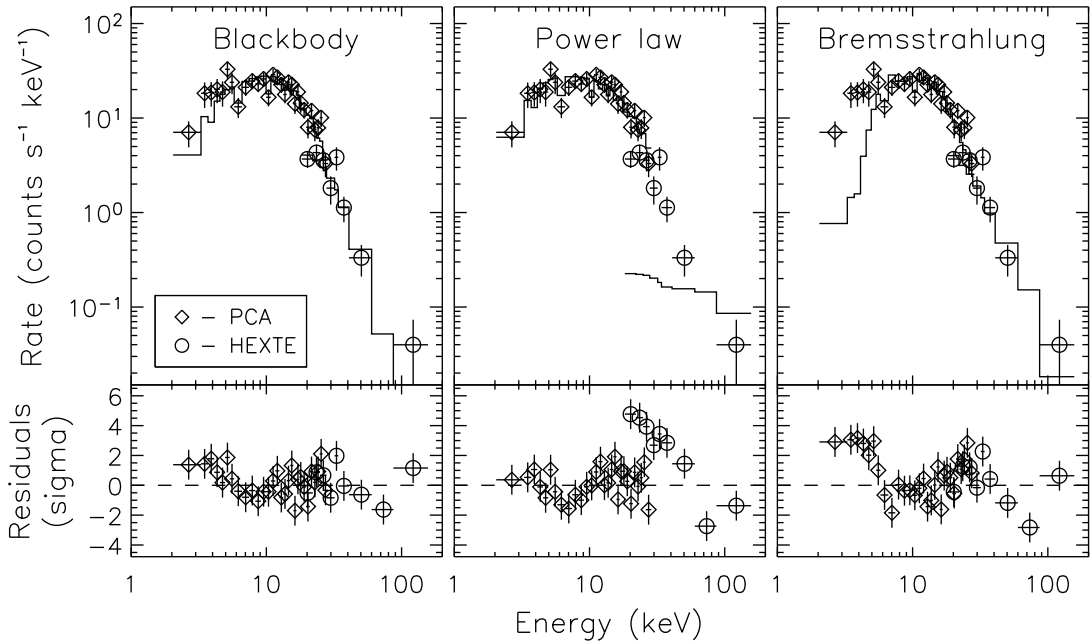,height=3.6in}}

\caption{Observed energy spectra for the PCA (diamonds) and HEXTE (circles) for
the primary peak of the 2004 February 16 burst (Burst 2) from \axpf.  The three
top panels show the data and the folded model count spectra for the blackbody,
power-law and bremsstrahlung models.  The bottom panels show the residuals of
the data to these model fits in units of standard deviations.  The spectral
fits are discussed in the text. \label{fig:burst_spectrum}}

\end{figure}

The PCA and HEXTE spectra for the brightest burst spikes (spike 1 of Bursts 1
and 2) were first grouped to include no fewer than 20 source counts per bin. 
Note that the background spectra were subtracted before grouping. In the PCA
data, there were $\sim$370 and 990 counts recorded (2$-$30 keV) during the
respective burst intervals.  For each burst, the spectra were read into {\tt
XSPEC} v11.2.0 and fit to three simple models -- power-law, bremsstrahlung, and
blackbody -- all modified by photoelectric absorption.  For the PCA and HEXTE
spectra, we simultaneously fit channels in the range 2.4$-$30 keV and 15$-$150
keV, respectively, allowing for separate normalizations for the two
instruments.  The power-law and bremsstrahlung models provided poor fits to the
spectrum of the primary peak of Burst 2 ($\chi^2$/dof = 125/34 and $\chi^2$/dof
= 98/34, respectively), whereas the blackbody model provided an acceptable fit
($\chi^2$/dof = 40/34).  All three model fits for this burst are shown in
Figure~\ref{fig:burst_spectrum}.  Similarly, the energy spectrum for the main
peak of Burst 1 was fit best with the blackbody model ($\chi^2$/dof = 16.1/17),
although the other two models provided statistically acceptable fits
($\chi^2$/dof = 22.6/17 for the power-law model and $\chi^2$/dof = 20.5/17 for
the bremsstrahlung model).  However, the measured column density was
unrealistically large at $\sim$80 $\times$ 10$^{22}$ cm$^{-2}$ for both the
power law and bremsstrahlung models.  The column density measured from the
persistent X-ray energy spectrum  is $N_{\rm H} = 1.1 \times 10^{22}$ cm$^{-2}$
(Gotthelf et al.\ 2004).  The blackbody model, on the other hand, returned an
upper limit for the column density ($N_{\rm H} < 2.0 \times 10^{22}$ cm$^{-2}$
at 90\% confidence) for Burst 2 consistent with the value measured from the
persistent X-ray emission.  Finally, we fit only the PCA spectra for these
bursts independently, and both the fit parameters and the quality of the fits
did not change significantly from the joint PCA/HEXTE fits.  We conclude that,
of the simple one-component models tested here, only the blackbody model
provides an adequate fit to the \axpf\ burst spike energy spectra.

\begin{figure}[!htb]

\vspace{-0.2in}

\centerline{
\psfig{file=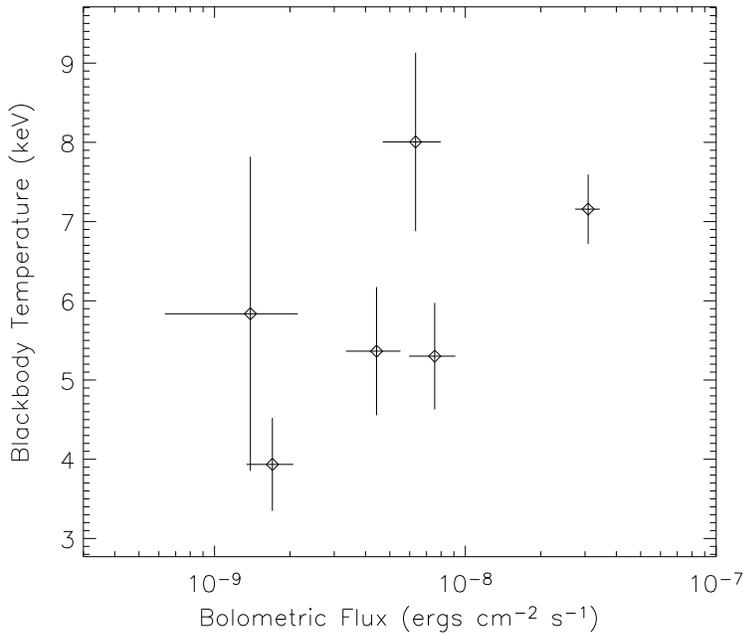,height=3.6in}}

\caption{Blackbody temperature ($kT$) versus bolometric flux for the six burst
spikes.  Note the increasing temperature with burst flux. The integration times
for the energy spectra used here vary from 0.5 to 2.0 s. 
\label{fig:bursts_bb_flux}}

\end{figure}

With the column density frozen at the value inferred from the persistent
emission spectral fits, we fit each of the six burst spike energy spectra (PCA
data only for the four remaining burst spikes) to the blackbody model.  Equally
good fits were obtained for the four additional burst spikes.  The blackbody
temperatures from all fits (including a fit to the primary peak of Burst 4) are
listed in Table~\ref{tab:bursts}.  None of the burst spike spectra show
evidence for discrete spectral features.  For our set of six burst spikes, we
find that the blackbody temperature is correlated with the flux of the burst
(Figure~\ref{fig:bursts_bb_flux}).  The probability that the measured spectral
temperatures are consistent with a constant temperature is 2 $\times$
10$^{-4}$.  The emitting radii range between 0.3 and 1.1 km (for a distance of
5 kpc [Gotthelf et al.\ 2004]), and show a marginal positive correlation with
burst flux (1 $\times$ 10$^{-3}$ chance probability).

Only the most energetic burst spike (the primary peak of Burst 2;
Figure~\ref{fig:burst2_peak1_zoom}) had a sufficient number of counts to search
for spectral evolution in time.  We split this burst ($-$0.15$-$2.0 s) into
three time intervals of $\sim$330 counts each.  The temperature does not change
significantly between the three intervals, although the limits on variability
are not very restrictive ($\lesssim$2 keV).

\subsection{Burst Tails}

Using the same procedure as described above, we extracted PCA spectra of the
tail emission, excluding the burst spikes, for the two brightest burst tails
(Bursts 1 and 2).  The remaining two burst tails did not have a sufficient
number of counts for meaningful spectral fits.  As with the burst spikes, we
fit the tail spectra to the blackbody, power-law and bremsstrahlung models, all
with interstellar absorption.  For these fits, the column density was frozen
at $N_{\rm H} = 1.1 \times 10^{22}$ cm$^{-2}$.  Only the blackbody model
provided an adequate fit ($\chi^2 =$ 74.5 for 63 dof) to the spectrum of the
X-ray tail of Burst 1, with a measured temperature of 2.6$\pm$0.2 keV.  The
power-law and bremsstrahlung models returned $\chi^2$/dof of 100/63 and 96/63,
respectively.  None of these simple models could adequately fit the tail
spectrum of Burst 2, although the blackbody model did provide the best fit. 
The $\chi^2$/dof for the three models were 55.9/29, 137/29, and 192/29,
respectively.

\begin{figure}[!htb]

\vspace{-0.2in}

\centerline{
\psfig{file=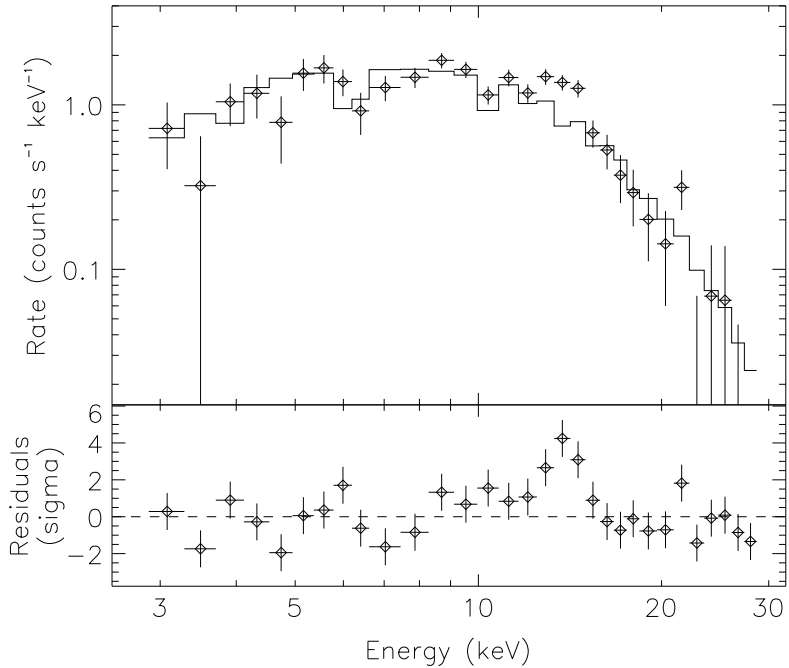,height=3.8in}}

\caption{{\it RXTE} PCA energy spectrum of the X-ray tail following the 2004
February 16 burst (Burst 2).  {\it Top} -- Observed count rates (diamonds) and
best-fit blackbody model (histogram) folded through the PCA response.  {\it
Bottom} -- Count rate residuals in units of sigma from the model.  Note the
excess near 13 keV.  See text for details.  \label{fig:burst2_tail_spec}}

\end{figure}

The residuals in the blackbody fit to the Burst 2 tail spectrum showed evidence
for an emission line at $\sim$13 keV consistent with the instrumental width of
the PCA (Figure~\ref{fig:burst2_tail_spec}).  When we included a narrow (i.e.\
zero width) Gaussian line in our model, the fit improved significantly ($\chi^2
=$ 30.0 for 27 degrees of freedom).  We estimated the probability of measuring
a change in $\chi^2$ this large (25.9) through a Monte Carlo simulation with
250,000 realizations of a blackbody spectrum.  Using {\tt XSPEC}, the spectra
were fit to a blackbody model and a blackbody plus a narrow line for energies
between 3 and 28 keV at 0.25 keV increments -- finer than the instrumental
resolution (1$-$2 keV).  For each simulated spectrum, the change in $\chi^2$
was measured between the pure blackbody and the blackbody plus line fit for the
line energy corresponding to the lowest $\chi^2$.  In {\it none} of the
simulated spectra was the improvement in $\chi^2$ larger than 24.4, thus, the
probability that a line of the strength measured would appear by chance is $<$4
$\times$ 10$^{-6}$.  The centroid of the line in the Burst 2 tail spectrum was
fit to a value 12.6$\pm$0.2 keV with an equivalent width of
1.43$^{+0.6}_{-0.5}$ keV.  The blackbody temperature for this fit was
3.6$\pm$0.2 keV.  The Burst 2 tail spectrum was of better statistical quality
than the Burst 1 tail spectrum, so the non-detection of a line in that spectrum
is not surprising.  The 3$\sigma$ upper limit on the equivalent width of a
narrow line at that energy for the Burst 1 tail spectrum is 4.1 keV.

We performed phase-resolved spectroscopy on each of these burst tails.  Using
spin ephemerides determined from phase-connected fits to longer stretches of
the persistent emission, we folded the data within each burst tail and
extracted PCA energy spectra with 8 phase bins for each.  For each tail, there
is no significant change of the temperature with pulse phase.  The limit on the
range of the blackbody temperature about the mean is $\pm$0.4 keV.

\begin{figure}[!htb]

\vspace{-0.4in}

\centerline{
\psfig{file=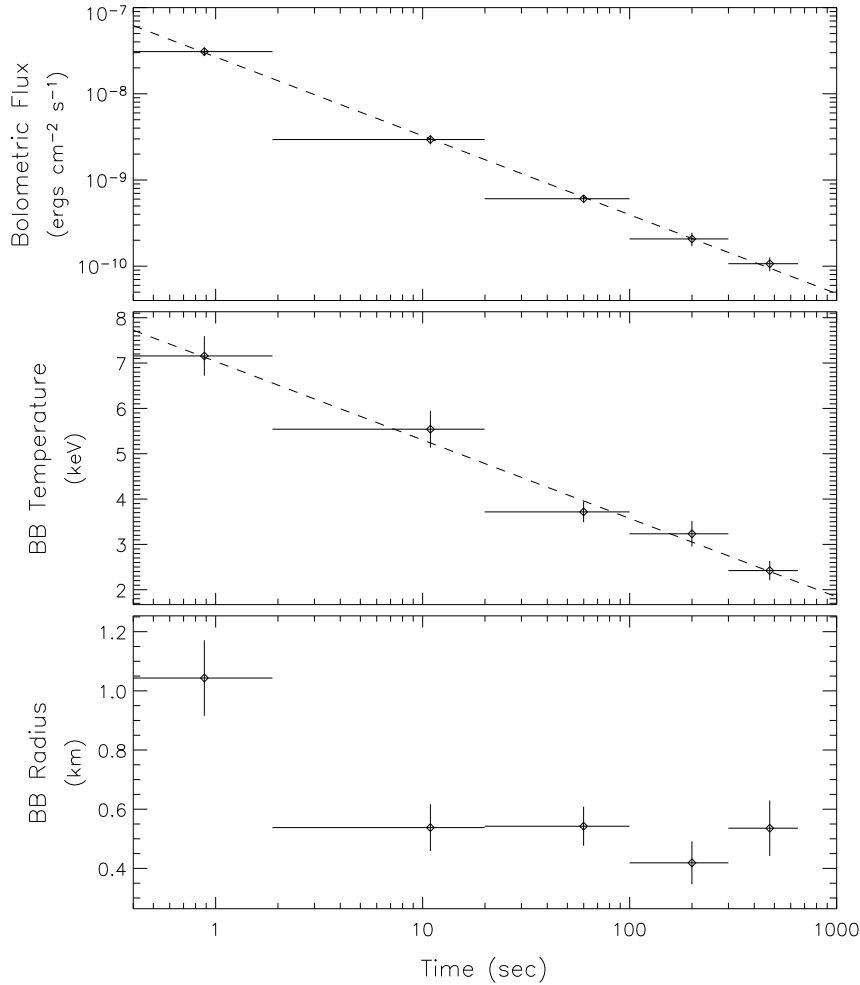,height=5.9in}}
\vspace{-0.2in}

\caption{Time evolution of the bolometric flux (top), blackbody temperature
(middle) and emitting radius (bottom) for the 2004 February 16 burst (Burst
2).  Times are relative to the burst peak.  Note the power-law decay ($F
\propto t^{-0.92}$) in the bolometric flux.  The blackbody temperature decays
approximately linearly in the given log-linear representation.  The inferred
emitting radius is 1 km (for an assumed distance of 5 kpc [Gotthelf et al.\
2004]) for the first 2 s, and is consistent with remaining constant at 0.5 km,
thereafter.  \label{fig:burst2_tail_bb}}

\end{figure}

Motivated by the observed dependence of the blackbody temperature on the burst
flux (Figure~\ref{fig:bursts_bb_flux}), we split the 2004 February 16 burst
(Burst 2 spikes and tail) into five segments of approximately equal source
counts.  We fit the individual spectra to the blackbody model and observe a
clear decay in the temperature as the flux during the burst/tail declines
(Figure~\ref{fig:burst2_tail_bb}).  The blackbody temperature decays linearly
with the logarithm of the time ($kT = B\log_{10}{(t - t_{\rm peak})} + kT_0$),
where $B$ = $-$1.73$\pm$0.16 keV $(\log_{10}{\rm s})^{-1}$.  The bolometric
flux decays more rapidly, following a power-law decay with an exponent
$-$0.92$\pm$0.02.  The isotropic energy emitted during the burst and tail is
8.7 $\times$ 10$^{38}$ ergs for an assumed distance of 5 kpc.  Note that there
is no sharp transition between the temperature during the primary burst peak
and the tail.  The blackbody emitting radius, however, drops from 1 km during
the burst to 0.5 km for the duration of the tail.

\section{Discussion}

\subsection{X-ray Tails}

After the peak of each burst from \axpf, the persistent (and pulsed) emission
increased, giving rise to extended X-ray tails.  There appears to be a causal
relationship between the bursts and tails in that each new burst spike elevates
the persistent flux.  For example, the persistent (and pulsed) flux of Burst 1
rose $\sim$50 s prior to the primary burst spike.  Following this spike, the
persistent flux rose even further and maintained this level for $\sim$60 s
through the next two (observable) burst spikes before slowly decaying away.  In
the tail following the most energetic burst (Burst 2), the data are consistent
with a constant emitting area whose surface cools as the bolometric flux
declines according to a power law in time.

Power-law flux decay following bursts has now been observed on numerous
occasions from four magnetar candidates (e.g.\ Woods et al.\ 2004).  As more
and more of these burst/tail pairs are recorded, their properties have been
found to be increasingly diverse.  The temporal decay indices of the X-ray
tails now range from $-$0.22 to $-$1 and the tail-to-burst energy ratios span
more than two orders of magnitude from 0.02 to $>$3.5.  Within the context of
the magnetar model, interpretations of these burst-induced transient flux
enhancements include cooling of the outer layers of a magnetar crust
(Lyubarsky, Eichler \& Thompson 2002) and enhanced magnetospheric emission due
to stronger currents driven by a twisting of magnetic field lines (Thompson et
al.\ 2000, 2002).  The exclusively thermal spectra of the X-ray tails observed
here suggests crustal cooling as opposed to optically-thin magnetospheric
emission.

Roughly 1\% of the total energy of each burst is expected to be conducted into
the crust and reradiated during the afterglow phase (Thompson \& Duncan 1995). 
In fact, the afterglow following the giant flare of 1998 August 27 from \sgrc\
comprised $\sim$2\% of the prompt burst energy (Woods et al.\ 2001), consistent
with the cooling of the outer layers of a magnetar crust (Lyubarsky et al.\
2002).  However, in Burst 2 from \axpf\ presented here, the tail energy
actually {\it exceeds} the burst energy by a factor $\sim$3.5.  Clearly, a
recycling of burst energy cannot explain the total tail emission seen here. 
This would suggest that the physical process which caused the short burst,
presumably a crustal fracture (see next section), also supplied the energy
required to power the subsequent X-ray tail.

The emission line in the energy spectrum of the X-ray tail of Burst 2 is
centered near $\sim$12.6 keV.  Emission lines at similar energies have been
detected from \sgrc\ (Strohmayer \& Ibrahim 2000) and \axpb\ (Gavriil et al.\
2002), although these features were detected during the prompt burst emission
at lower confidence levels.  It has been suggested that in a magnetar, proton
cyclotron emission lines can be excited following bursts with sufficient
radiative pressure to drive a thin layer of particles off the stellar surface
(Ibrahim et al.\ 2001).  The detection of the line {\it following} this burst
from \axpf\ is consistent with this picture.  If this line represents the
fundamental frequency for protons, the required local field would be 2 $\times$
10$^{15}$ G, or $\sim$7 times higher than the dipole field strength inferred
from the spin down parameters (Ibrahim et al.\ 2004).  Note that local magnetic
fields higher than the dipole field are possible and furthermore, there are
significant uncertainties in the magnetic field calculation for both the proton
cyclotron energy and dipole spin down (e.g.\ Thompson et al.\ 2000).

\subsection{Two Classes of Magnetar Bursts}

A new paradigm appears to be emerging when we add the \axpf\ bursts to the
existing magnetar candidate burst sample.  We now are able to identify a class
of bursts that ($i$) have thermal energy spectra ($ii$) are correlated with the
pulsed flux maxima and ($iii$) exhibit extended X-ray tails lasting tens to
hundreds of seconds.  All four of the bursts from \axpf\ reported here and
several others from both \axpb\ (Gavriil et al.\ 2002) and \axpa\ (Kaspi et
al.\ 2003; Gavriil et al.\ 2004) seem to be part of this class.  One of the two
bursts detected from \axpb\ in 2001 (Gavriil et al.\ 2002) had a significant
tail of emission lasting $\sim$50 s.  While the full duration of this burst was
long, it started with a brief ($\sim$1 s), bright spike that had a short rise
time (20 ms), very similar to the morphology of the burst shown in
Figure~\ref{fig:burst2_peak1_zoom}.  The other \axpb\ burst was similar, but
there was no extended X-ray tail detected, although this could simply be due to
insufficient sensitivity.  The peaks of these two bursts occurred near pulse
maximum, similar to the \axpf\ bursts.  Another burst recorded from \axpb\ in
2004 (Kaspi et al.\ 2004) exhibits a brighter X-ray tail (Gavriil et al.\ 2005)
than either of the 2001 bursts from that source.  The properties of this tail
are similar in many respects to the X-ray tail following Burst 2 from \axpf.

\begin{figure}[!p]

\centerline{
\psfig{file=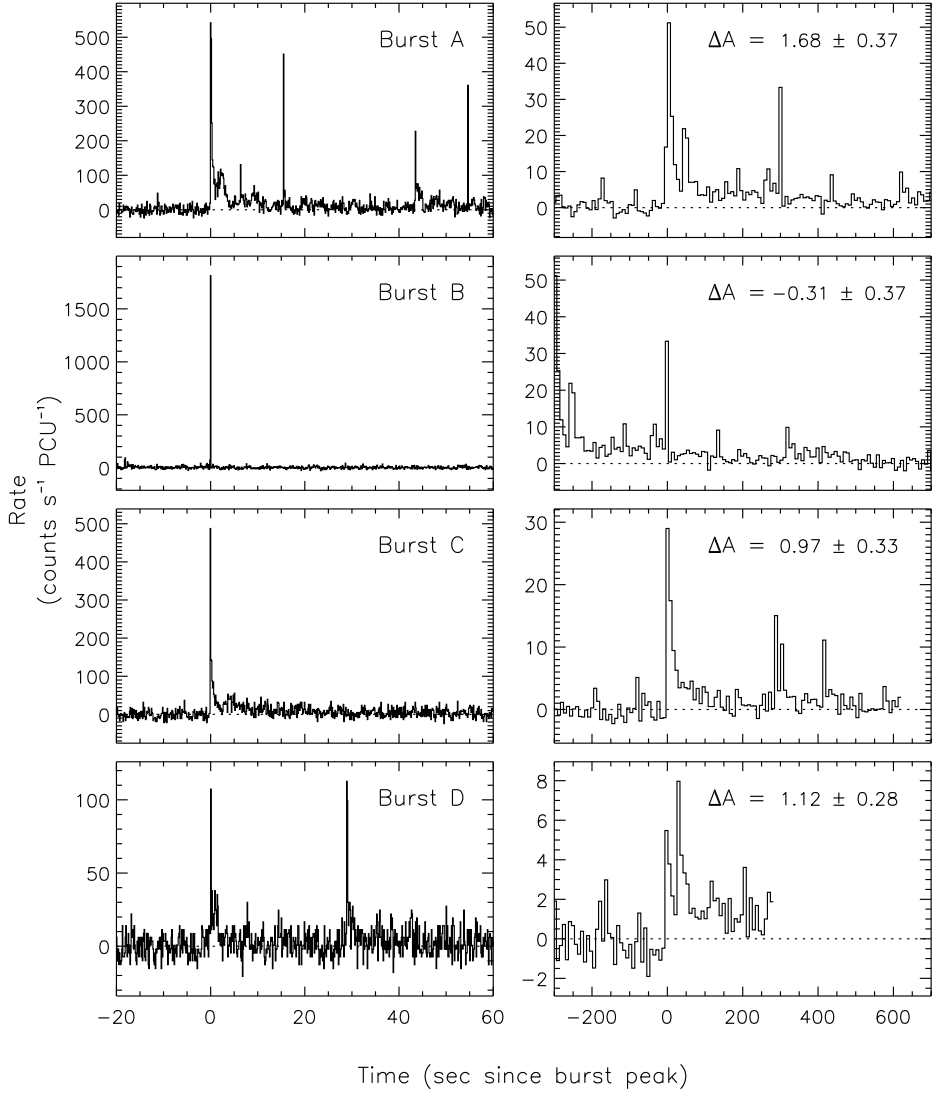,height=6.5in}}

\caption{Time histories of four selected bursts recorded from \axpa\ with the
{\it RXTE} PCA (2.1$-$30.0 keV).  The four left panels show each burst with
0.125 s time resolution.  Right panels show the same bursts over an expanded
time range with 8 s time resolution.  All light curves are background
subtracted where both the instrumental background (PCABACKEST) and power-law
flux decay intrinsic to the source (Woods et al.\ 2004) have been removed.  The
parameter ($\Delta A$) printed in the upper right corner of each expanded light
curve is the change in pulsed amplitude (2$-$30 keV) for 200 s integrations
prior to and following each burst.  The units of $\Delta A$ are counts s$^{-1}$
PCU$^{-1}$. \label{fig:bursts_2259_z2m}}

\end{figure}

The 2002 June outburst from \axpa\ contained $\sim$80 individual bursts within
a single 11-ks \rxte\ observation (Gavriil et al.\ 2004).  The arrival time
distribution of these bursts in rotational phase of the pulsar, like those from
\axpb\ and \axpf, was correlated with the pulse maxima.  The \axpa\ bursts also
showed a hardness-fluence correlation qualitatively similar to what we find
between the burst flux and blackbody temperature of \axpf\ bursts.  Several of
the bursts recorded from \axpa\ were brief ($\sim$0.1 s duration) spikes
similar to those detected from SGRs ({G\"o\u{g}\"u\c{s}} et al.\ 1999, 2000,
2001).  However, a fraction of the \axpa\ bursts clearly had extended X-ray
tails lasting hundreds of seconds in some cases.  The high concentration of
burst activity and the rapidly decaying X-ray flux upon which these bursts were
superposed (Woods et al.\ 2004) made it difficult to identify these tails. 
Three examples of bursts with extended tails from \axpa\ are shown in
Figure~\ref{fig:bursts_2259_z2m} (Bursts A, C and D) plus one event that does
not show a post-burst flux enhancement (Burst B).  Moreover, the bursts from
\axpa\ with X-ray tails also exhibited a significant increase in pulsed
amplitude following the burst.  The change in 2$-$30 keV pulsed amplitude
($\Delta A$) for 200 s integrations prior to and following each burst is
printed on the right-most panels of Figure~\ref{fig:bursts_2259_z2m}.  The
pulsed amplitude of \axpa\ increases significantly ($\gtrsim$3$\sigma$)
following Bursts A, C and D, but shows no significant change following Burst
B.  Interestingly, Burst B also has the highest peak flux (on time scales
shorter than 1 s) of all bursts detected from \axpa; thus bursts with high peak
flux are not necessarily accompanied by X-ray tails or increases in pulsed
flux.

The bursts we have observed from these three AXPs differ from SGR bursts in a
few key ways.  First, hundreds of bursts from \sgrc\ recorded in 1998 were
consistent with being distributed randomly in pulse phase (Palmer 2002),
whereas all three bursting AXPs have shown a correlation with the pulsed X-ray
flux.  Second, the ratio of burst-to-tail energy in AXP is very different from
that of SGR bursts.  X-ray tails from SGRs have been detected, however, only
following high-fluence SGR bursts (Woods et al.\ 2001; Ibrahim et al.\ 2001;
Lenters et al.\ 2003; Feroci et al.\ 2003) where the energy during the burst
interval is many times larger than the energy emitted during the tail (Lenters
et al.\ 2003).  For the bursts with X-ray tails seen from the AXPs, the tail
energy is larger than the energy emitted during the primary burst peak.  Third,
extended X-ray tails are a much rarer phenomenon following SGR bursts, in spite
of the fact that many SGR bursts are orders of magnitude brighter than the
bursts seen from these three AXPs.  Fourth, the SGR bursts with X-ray tails
exhibit a distinct morphology showing an abrupt flux termination that nominally
marks the transition point between burst and tail (Woods \& Thompson 2004).  In
a single case where there have been sensitive observations during one of these
burst-to-tail transition points, we observed a spectral discontinuity
coincident with the sudden drop in burst flux (Ibrahim et al.\ 2001).  The AXP
bursts show a quasi-exponential decay with no sharp transition point between
burst and tail either in spectrum or morphology.  Finally, the broad-band
spectrum (2$-$150 keV) of bursts from \sgrc\ has recently been shown to be well
modeled by the sum of two blackbodies (Feroci et al.\ 2004; Olive et al.\
2004).  At least for \axpf, the spectrum is well fit by a single blackbody over
a similar energy range.

It is clear that this new type of burst is more common in AXP sources as
compared to the SGRs.  It should be noted that at least \axpa\ emits both types
of bursts, so the two burst types are apparently not mutually exclusive from
source to source or even within a given burst active period.  For the purposes
of the following discussion, we will refer to the traditional SGR bursts as
Type A bursts and this potentially different type of bursts seen in AXPs as
Type B bursts.

\subsection{Magnetar Burst Models}

The Type B bursts we have now observed from three different sources have a set
of common characteristics that place constraints on their origin and geometry. 
In particular, the correlation of the burst arrival times with the pulsed flux
maxima suggests some degree of collimation.  Based upon the average phase
offset from the peak (0.2 cycles), we estimate the opening angle of the
emission cone to be $\sim$70$^{\circ}$.  In all likelihood, the bursts are
emitted at irregular intervals, yet we only observe those whose cone of
emission includes our line of sight, thus producing the temporal correlation
with pulse maximum.  The observed properties of the \axpf\ bursts suggest that
they originate from the same vicinity which produces the pulsed maximum,
presumably at or above the polar cap.  Given the thermal shape of these burst
spectra and their absence during pulse minimum, the bursts are probably
produced at, or very near the stellar surface.  Note that this does not require
the persistent X-ray pulsations be produced at the surface (i.e.\ the
pulsations may originate from higher up in the magnetosphere above the polar
cap).

Collimation of the burst flux suggests that one would expect to observe
``naked'' X-ray tails or enhanced persistent emission without observable
bursts.  Burst 4 and the first 50 s of Burst 1 are examples of such events. 
Each cycle of these burst intervals, including the primary peak of Burst 4, has
a morphology consistent with the pulse shape (Figure~\ref{fig:peak_pulse}).  In
these instances, a burst may have been emitted while the polar cap was pointed
away from our line of sight, so we only detected the X-ray tail of the burst,
but not the burst itself.  The fraction of observed naked tails to tails
accompanied by bursts is likely to be small for several reasons.  First, there
is an observational selection effect against the detection of naked tails.  The
pulsed flux enhancements are simply more difficult to detect than the bursts
due to the relatively modest change in pulsed amplitude following bursts and
the long integration times (150$-$300 s) required to detect this change.  Since
the pulsed flux enhancements decay away rapidly, long integration times
effectively dilute the signal.  Second, the collimation of emission from the
stellar surface will be reduced by the strong gravitational potential well of
the neutron star, thus revealing a larger fraction of the neutron star surface
to our line of sight (e.g.\ Psaltis, \"Ozel \& DeDeo 2000).  Third, the
durations of the bursts (0.5$-$2 s) are a significant fraction of the pulse
period; thus one is likely to detect the burst even if it starts while the pole
is pointed away from our line of sight.  Finally, multiple bursts are often
times emitted during each burst event, so with each burst the chances that one
or more will cross our line of sight is increased.

What then do these constraints say about the physical mechanism that produces
the bursts?  Within the context of the magnetar model, two physical triggers
have been proposed to explain magnetar bursts: crust fracturing (Thompson \&
Duncan 1995) and magnetospheric reconnection (Lyutikov 2002).  Note that a
localized crustal fracture may also induce sudden magnetospheric reconnection
within a narrow bundle of field lines anchored to the crust (Thompson \& Duncan
2001).  In the crust fracturing model, two distinct mechanisms for converting
the stored potential energy in the crust to observable X-rays have been put
forth (Thompson \& Duncan 2001; Heyl \& Hernquist 2004).  However, the end
product of an expanding pair-photon plasma that emits predominantly blackbody
emission is the same for each mechanism.  Thus the two are, in the present
understanding, indistinguishable observationally.  The properties of the bursts
resulting from the two physical triggers, on the other hand, are expected to
have observational differences (e.g.\ Lyutikov 2002).

The qualitative differences in the two models would favor the magnetospheric
reconnection trigger for Type A bursts and crust fracturing for Type B bursts. 
A magnetospheric origin would lend itself to more isotropic emission having no
preference for a particular pulse phase.  Alternatively, the crust fracture
model would naturally produce a phase dependence of the burst emission for a
localized active region on the crust.  Moreover, the tendency of the bursts to
occur near pulse maximum is consistent with the strain in the crust being
highest in regions where the field is strongest (Thompson \& Duncan 1995) -- at
the polar caps.  Lyutikov (2002) argued that the conditions most favorable for
magnetic reconnection are present in magnetars with magnetospheres having
strong electrical currents and a high degree of magnetospheric shear. 
Thompson, Lyutikov \& Kulkarni (2002) have argued that the SGRs, with their
strong non-thermal X-ray spectral components, are most likely to match these
conditions, thus making SGRs prime candidates for magnetospheric reconnection. 
Finally, Lyutikov (2002) shows that the magnetospheric reconnection model can
account for the observed hardness-fluence anti-correlation seen in bursts from
\sgra\ ({G\"o\u{g}\"u\c{s}} et al.\ 2001).  The bursts from the AXP \axpa\
exhibit a positive correlation (Gavriil et al.\ 2004).

At this time, there are very few quantitative model predictions for either the
crustal fracture model or the magnetospheric reconnection model, particularly
in terms of anticipated energy spectra and light curves.  One notable exception
is the prediction of bright radio bursts accompanying X-ray bursts triggered by
magnetospheric reconnection (Lyutikov 2002).  The lack of detailed model
predictions limits the rigor with which we can compare and contrast these two
models with the available data.  In addition, more work is needed in analyzing
existing data sets.  For example, is the apparent spectral difference, one
blackbody in Type B bursts versus two blackbodies in Type A bursts, ubiquitous
among these different burst sources?  Is there a small, as yet undiscovered,
fraction of Type B bursts hidden in the SGR burst database?  Answering these
questions will help illuminate the full scope of magnetar candidate burst
properties and ultimately improve our physical understanding of the burst
mechanism.

\acknowledgments{\noindent {\it Acknowledgements} --  }  PMW thanks Chris
Thompson, Sandy Patel, Ersin {G\"o\u{g}\"u\c{s}}, Allyn Tennant, Keith Arnaud,
and Joseph Ventura for useful discussions.  The authors thank the anonymous
referee for carefully reading the manuscript and for providing useful
suggestions on how to improve it.  PMW is grateful for support from NASA
through grant NNG~04GB20G.  VMK acknowledges funding from  Science and
Engineering Research Canada, Canadian Institute for Advanced Research, and
Fonds Quebecois de la Recherche sur la Nature et  les Technologies.

\end{document}